\documentstyle[epsfig,aps,multicol]{revtex}
\begin{document}
\title{Landau Levels from the Bethe Ansatz Equations}
\draft 
\author{K. Hoshi$^1$ and Y. Hatsugai$^{1,2,*}$}
\address{$^1$Department of Applied Physics, University of Tokyo,
7-3-1 Hongo Bunkyo-ku, Tokyo 113-8656, Japan\\
and\\
$^2$PRESTO, JST, Japan}
\date{July 29,1999}
\maketitle
\begin{abstract}
The Bethe ansatz (BA) equations for the two-dimensional Bloch electrons 
in a uniform magnetic field
are treated in the weak field limit.
We have calculated energies near the lower boundary of the energy spectrum
up to the first nontrivial order.
It corresponds to calculating a finite size correction
for the excitation energies of the BA solvable lattice models
and gives the Landau levels in the present problem.
\end{abstract}

\begin{multicols}{2}

The system of two-dimensional Bloch electrons
in a magnetic field has very rich structure.
The ratio of a lattice constant and a magnetic length,
which are two fundamental length scales, is crucially important.
The incommensurability brings fractal structures  in the problem which
can be observed in the energy spectrum
and the wavefunctions \cite{hof}.
Famous Hofstadter's butterfly diagram is a typical demonstration of the
structure.
It is fractal and has a self-similar structure.
It also has an interesting relation to
the quasiperiodic systems (quasicrystals) \cite{hir}.

Recently intrinsic importance of a quantum-mechanical phase
is stressed as a geometrical phase, 
which is essential in the Aharonov-Bohm effect,
the Berry's phase, anyons and the quantum Hall effect \cite{geo}.
The Bloch electrons in a magnetic field
also have importance 
as a stage of a typical realization of the geometrical phase.
It was first noticed by Zak in the study of a magnetic translation
group \cite{zak}.
The effect of the geometrical phase
for the Bloch electrons was
pursued in the studies of the quantum Hall effect
and fundamental relations between
the Hall conductance and several topological invariants
were discovered (the Chern number \cite{tho}
and the winding number on the complex energy surface
\cite{hat1}).
This is a geometrical aspect of the problem.

For an algebraic property of the problem which
originates from the magnetic translation group,
there was a breakthrough recently.
Wiegmann and Zabrodin found a relation 
between the Hamiltonian of the Bloch electrons in a magnetic field
and the quantum group $U_q(sl_2)$ \cite{wie,fad}.
Using the relation, 
the Schr\"odinger equation
is rewritten as a functional equation 
and the Bethe ansatz (BA) equations are derived.
The quantum group is a relatively new mathematical concept.
It is a kind of generalization of the Lie algebra
which is characterized by the so-called
$q$-parameter.
This $q$-parameter represents the non-commutativity of the
elements in the quantum group which
is related to the non-commutativity of the magnetic translation group.
Mathematically rich structures of the quantum group enable
us to obtain new insights of the physical problem.
Since the BA equations are high degree multi-variable
ones, it is difficult to obtain an explicit solution.
It usually occurs for the BA equations 
of the BA solvable lattice models
(the XXZ chain and the Hubbard chain).
However, for the present problem, the explicit solution of
the BA equations for the zero energy was obtained
by Hatsugai-Kohmoto-Wu \cite{hat2,hat3}.
When the flux per plaquette is irrational,
the distribution function
of the BA roots is nowhere differentiable
and the polynomial representation of the $U_q(sl_2)$ gives
a polynomial with quasiperiodic coefficients.
It implies that the wavefunction has a multifractal character.
Although this explicit solution is only
restricted to the zero energy, 
the other energies were also studied numerically \cite{hat3}.
Abanov, Talstra and Wiegmann gave beautiful results
based on the string type ansatz when
the flux per plaquette is an irrational golden mean \cite{aba}.
Krasovsky derived an integral equation for the distribution function
of the BA roots
in the weak field limit
and calculated the energy
of the lowest energy band \cite{kra}.
This corresponds to calculating a finite size correction of the
ground state energy in the usual BA solvable models.

In this letter, we focus on the weak field limit $\phi\to0$
where the flux per plaquette is given by $\phi=Ba^2/\Phi_0$
in units of a flux quantum $\Phi_0=hc/|e|$
($a$ is a lattice constant and $B$ is a magnetic field).
Physically this weak field limit can be
understood as a continuum limit from the lattice system
where the lattice spacing $a$ vanishes
with the magnetic field $B$ fixed.
We calculate energies of the lowest several states analytically
up to the lowest nontrivial order in $\phi$.
It corresponds to calculating finite size corrections for excited
state energies of the usual BA solvable lattice models.
Mathematically this weak field limit is a semiclassical limit
where the classical $sl_2$ is recovered
from the quantum $U_q(sl_2)$.

\begin{center}
\begin{figure}
\epsfig{figure=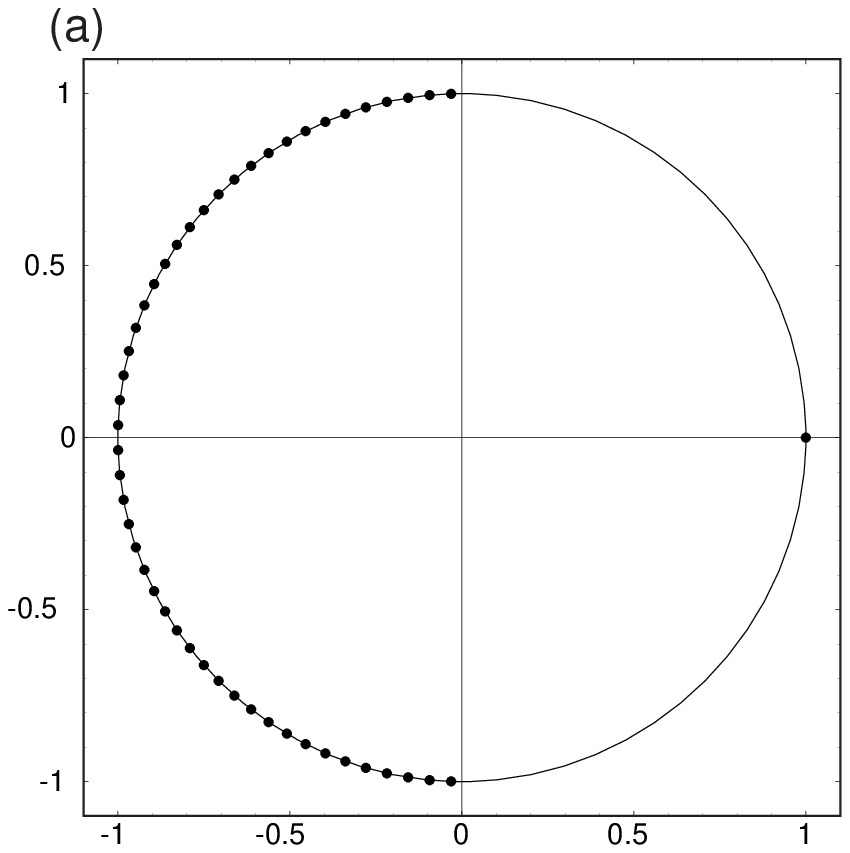,width=8.2cm}
\epsfig{figure=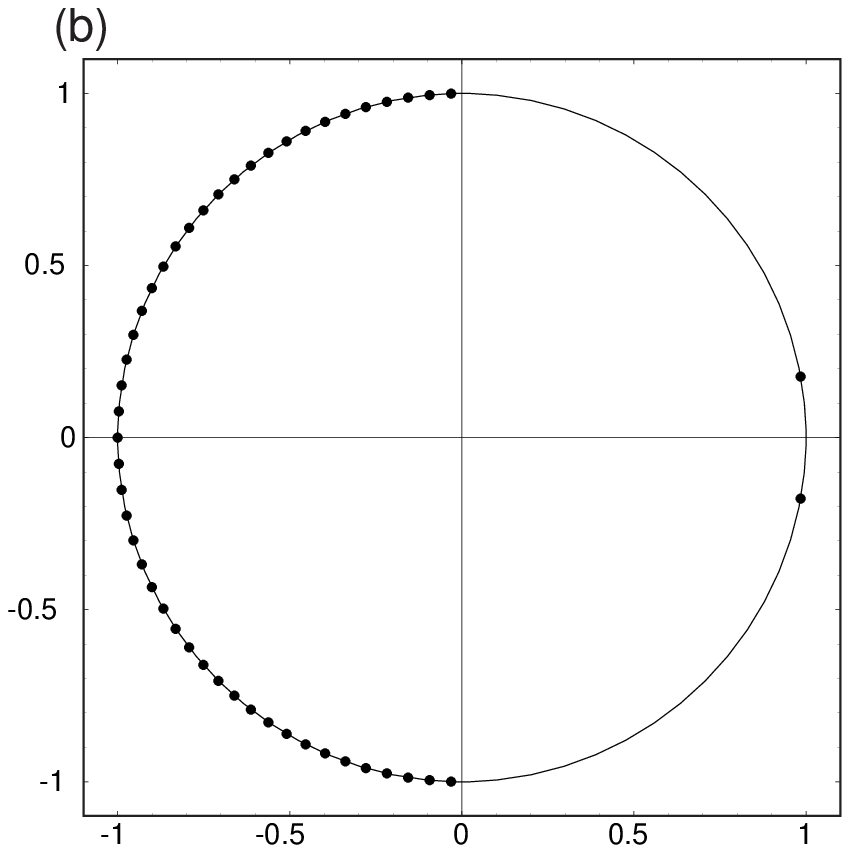,width=8.2cm}
\begin{minipage}{8.5cm}
\caption{\label{fig:roo} Numerical solutions for the roots of the Bethe ansatz equations: 
(a) for the second lowest energy band
and (b) for the third band ($\phi=1/50$).
}
\end{minipage}
\end{figure}
\end{center}

The Hamiltonian of electrons on a square lattice
in a uniform magnetic field
is given by
\begin{eqnarray}
H=T_x+T_y+\mbox{h.c.},\nonumber
\end{eqnarray}
where $T_x$ and $T_y$ are the covariant translation operators
$T_x=\sum_{m,n}c_{m+1,n}^{\dagger}e^{i\theta_{m,n}^x}c_{m,n}$,
$T_y=\sum_{m,n}c_{m,n+1}^{\dagger}e^{i\theta_{m,n}^y}c_{m,n}$,
and $c_{m,n}$ is an annihilation operator for an electron
at a site ($m,n$).
Discrete rotation of the phases $\theta_{m,n}^x$ and $\theta_{m,n}^y$
gives a flux per plaquette $\phi$
: $\sum_{\mbox{{\tiny plaquette}}}\theta_{m,n}=
\theta_{m,n}^x+\theta_{m+1,n}^y-\theta_{m,n+1}^x-\theta_{m,n}^y
=-2\pi\phi$.
In the diagonal gauge ($\theta_{m,n}^x=\pi(m+n)\phi$,
$\theta_{m,n}^y=-\pi(m+n+1)\phi$),
the Hamiltonian in the momentum space is given as
$H=\sum_{\bf k}H({\bf k})$, where
$H({\bf  k})
=(e^{-ik_x}X+e^{ik_y}X^{-1})Y+Y^{-1}(e^{-ik_y}X+e^{ik_x}X^{-1})$
with $2Q\times2Q$ matrices $X$ and $Y$
($X_{ij}=1$ if $i-j\equiv1\pmod{2Q}$, $X_{ij}=0$ otherwise,
$Y=\mbox{diag}(q,q^2,\ldots,q^{2Q})$).
When one takes $P$ to be odd and chooses momenta $(k_x,k_y)$ on the
so-called midband line
($k_+=(k_x+k_y)/2\equiv\pi/2\pmod{\pi/Q}$),
the Hamiltonian is given by
\begin{eqnarray}
H_{\tiny \mbox{midband}}
=i(q-q^{-1})\{\rho_c(B)+\rho_c(C)\},\nonumber
\end{eqnarray}
where $\rho_c(B)$ and $\rho_c(C)$ 
supplemented by $\rho_c(A)$, $\rho_c(D)$ 
are the cyclic representations
of the quantum group $U_q(sl_2)=\{A,B,C,D\}$
\cite{wie,hat3}.
Here the quantum group enters in the problem
where the $q$-parameter is given by
$q=e^{i\pi \phi}$.
Since we consider the case $\phi=P/Q$ 
with mutually prime integers $P$ and $Q$,
it implies that the $q$ is a root of unity ($q^{2Q}=1$).
This cyclic representation is only possible
for the $q$ which is a root of unity.
The classical $sl_2$ is a spin algebra
and it has a usual highest weight representation
given by the differential operators.
Correspondingly the quantum group
$U_q(sl_2) $ also has a highest weight representation
given by the difference operators.
The highest weight representation implies that
finite degree polynomials can span the bases.
On the other hand,
the cyclic representation does not have a classical correspondence.
However,
using a relation between the
highest weight representation and the cyclic representation,
the discrete Schr\"odinger equations of the
Bloch electron in a magnetic field
are extended
to a functional equation
for a polynomial $\Psi(z)=\prod_{k=1}^{Q-1}(z-z_k)$
of the degree $Q-1$ \cite{wie} as
\begin{eqnarray}
i(z^{-1}+qz)\Psi(qz)-i(z^{-1}+q^{-1}z)\Psi(q^{-1}z)=\varepsilon\Psi(z).
\label{feq}
\end{eqnarray}
Putting $z=z_m$ and also comparing the coefficients of the highest order term,
we have the BA type equations \cite{wie} as
\begin{eqnarray}
\frac{z_m^2+q}{qz_m^2+1}=q^Q\prod_{k=1}^{Q-1}\frac{qz_m-z_k}{z_m-qz_k}
,\quad m=1,2,\ldots,Q-1,\cr
\varepsilon=iq^Q(q-q^{-1})\sum_{k=1}^{Q-1}z_k.\label{BAE1}
\end{eqnarray}
For the semiclassical case
$\phi=1/Q$
which we consider in the paper, 
all the BA roots are on a unit circle,
$|z_k|=1$ ($k=1,2,\ldots,Q-1$) \cite{hat3} (see Fig.\ref{fig:roo}).

Therefore, we set $z_k=e^{i\varphi_k}\;(\varphi_k\in{\bf R})$.
Then the BA equations (\ref{BAE1}) are written as
\begin{eqnarray}
\frac{\cos(\varphi_m-\frac{\pi}{2Q})}{\cos(\varphi_m+\frac{\pi}{2Q})}
&=&-\prod_{k=1}^{Q-1}
\frac{\sin\frac{1}{2}(\varphi_m-\varphi_k+\frac{\pi}{Q})}
{\sin\frac{1}{2}(\varphi_m-\varphi_k-\frac{\pi}{Q})}.\label{BAE2_1}
\end{eqnarray}
In the limit $Q\to\infty$, the distribution function $\rho(\varphi)$
of the BA roots is well defined and
the energy is given by
\begin{eqnarray}
\varepsilon
&=&2\pi\int_{-\pi}^{\pi}e^{i\omega}\rho(\omega)d\omega,\nonumber
\end{eqnarray}
where $\rho(\varphi_k)=\lim_{Q\to\infty}1/Q\delta_k$
and $\delta_k=\varphi_{k+1}-\varphi_k$.
Recently, Krasovsky calculated
the energy of the lowest band up to the $1/Q$ order
and an integral equation for $\rho$
was derived \cite{kra}.
Extending the method in Ref.\cite{kra},
we calculate energies of the general bands
in the vicinity of the lower boundary of the spectrum
($\varepsilon =-4+O(1/Q)$).
To obtain the energies, one needs the distribution of the BA roots.
For the $n$-th lowest energy band with $n=O(1)$,
$Q-n$ roots are almost uniformly distributed 
on the left semicircle with a unit radius
and the other $n-1$ roots are 
on the right semicircle \cite{hat3}
(see Fig.\ref{fig:roo}).
We determine the arguments of the $n-1$ roots 
up to the order of unity.
We denote the arguments of the $Q-n$ roots and those of the other $n-1$ roots 
as $\theta$ and $\tau$ respectively
($\theta\in(\pi/2,3\pi/2)$ 
and $\tau\in(-\pi/2,\pi/2)$).
In general, the $n-1$ roots consist 
of clusters separated by finite distances.
The roots within each cluster
converge to the single point in the $Q\to\infty$ limit.
Then the arguments of the BA roots are written as
$\{\tau_{11},\ldots,\tau_{1n_1}\},
\{\tau_{21},\ldots,\tau_{2n_2}\},\ldots$
where $\tau_{1i}\to c_1$,
$\tau _{2j}\to c_2,\ldots$ ($Q\to\infty$)
($c_i\neq c_j$, $i\neq j$).
Further we assume $c_i\neq\pm\frac{\pi}{2}$ (see Fig.\ref{fig:roo}).
Then the BA equation (\ref{BAE2_1})
for $\tau_{11}$
in the limit $Q\to\infty$ gives
\begin{eqnarray}
0&=&\lim_{Q\to\infty}\sum_k
\ln\left|
\frac{\sin\frac{1}{2}(c_1-\theta_k+\frac{\pi}{Q})}
{\sin\frac{1}{2}(c_1-\theta_k-\frac{\pi}{Q})}\right|\cr
&&+\lim_{Q\to\infty}\sum_{i=2}^{n_1}
\ln\left|
\frac{\sin\frac{1}{2}(\tau_{11}-\tau_{1i}+\frac{\pi}{Q})}
{\sin\frac{1}{2}(\tau_{11}-\tau_{1i}-\frac{\pi}{Q})}\right|.
\end{eqnarray}
The first term is estimated
as $\ln\frac{1-\sin c_1}{1+\sin c_1}$.
Adding the BA equations for the roots
$\{\tau_{11},\ldots,\tau_{1n_1}\}$, we have
$n_1\ln\frac{1-\sin c_1}{1+\sin c_1}=0$.
It means $c_1=0$.
Similarly, we have $c_2=\cdots=0$. 
Therefore, there is only one cluster on the right semicircle (see
Fig.\ref{fig:roo}) and
\begin{eqnarray}
\tau_i\to 0\quad(Q\to\infty).\nonumber
\end{eqnarray}

\begin{center}
\begin{figure}
\epsfig{figure=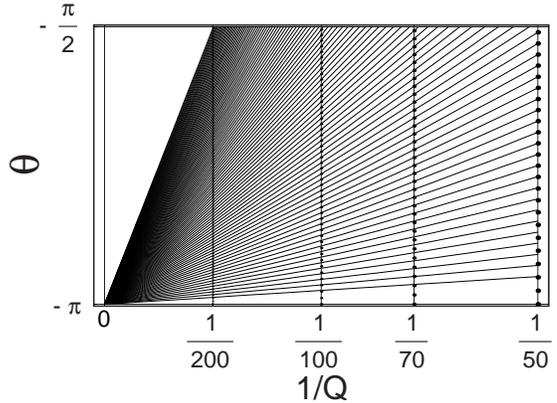,width=9cm}
\begin{minipage}{8.5cm}
\caption{\label{fig:iqr} Numerical solutions for the arguments of the roots $\theta_k$
for the third lowest band
($Q=$50, 70, 100 and 200).
The straight lines are the asymptotic behavior for $Q\to\infty$,
$\theta_k=-\pi+(k+3/2)\pi/Q$ $(k=1,2,\ldots,(Q-4)/2)$.
}
\end{minipage}
\end{figure}
\end{center}

Now let us calculate the rest of the BA roots $\theta_i$
and the energy for the ($2p+1$)-th lowest energy band
(we assume that the $Q$ is even for definiteness).
By the symmetry of the BA roots distribution \cite{hat2,hat3,kra}
(see Fig.\ref{fig:roo}),
we write arguments of the BA roots as
$\theta_0=-\pi$, $\theta_k
=-\pi+\frac{\pi k}{Q}+\sum_{j=0}^{k-1}\Delta_j$,
$\theta_{-k}=-\theta_k$,
$k=1,2,\ldots, \frac{Q}{2}-p-1$
where $\Delta_j=\theta_{j+1}-\theta_j-\frac{\pi}{Q}$.
Note that $\sum_{j=0}^{k-1}\Delta_j$ vanishes in the 
$Q\to\infty$ limit,
since $\Delta_j=o(\frac{1}{Q})$.
Moreover, $\Delta_k$ is exponentially small 
for the $k$ of the order $Q$ ($k\sim Q$) \cite{del}.
Therefore, we set 
$\theta_k=-\pi+\frac{\pi k}{Q}+\frac{s}{Q^{\delta}}+o(\frac{1}{Q^{\delta}})$ 
for $k\sim Q$
where $s$ and $\delta$ are constants independent of $k$ and $Q$.
Numerical solutions for $\theta_k$ are shown in Fig.\ref{fig:iqr}.
For the energy $\varepsilon$, 
we have $\varepsilon=-4+\frac{4 \pi p}{Q}+\frac{4s}{Q^{\delta}}
+o(\frac{1}{Q},\frac{1}{Q^{\delta}})$
from Eq.(\ref{BAE1}).
To fix the energy, we calculate $s$ and $\delta$.
Now let us rewrite the BA equations (\ref{BAE2_1}) 
by $\theta_k$ and $\tau_l$ as
\begin{eqnarray}
\ln\left|
\frac{\cos(\theta_m-\frac{\pi}{2Q})}{\cos(\theta_m+\frac{\pi}{2Q})}\right|
&=&\sum_k\ln\left|
\frac{\sin\frac{1}{2}(\theta_m-\theta_k+\frac{\pi}{Q})}
{\sin\frac{1}{2}(\theta_m-\theta_k-\frac{\pi}{Q})}\right|\cr
&&+\sum_l\ln\left|
\frac{\sin\frac{1}{2}(\theta_m-\tau_l+\frac{\pi}{Q})}
{\sin\frac{1}{2}(\theta_m-\tau_l-\frac{\pi}{Q})}\right|.\nonumber
\end{eqnarray}
Note that 
if the first summation is naively estimated as integral, 
the integrand diverges at the points
$|m-k|\ll Q$ ($x\ll y$ means $\frac{x}{y}\to0$ ($Q\to\infty$)). 
In order to avoid this, we divide the summation into two
parts $S$ and $R$,
where $S$ is the contribution from $k=m-N,\ldots,m+N$ ($1\ll N\ll Q$)
and $R$ is that from the rest $k$
(we take $m$ such that $\theta_m\not\to\pm\frac{\pi}{2}$ ($Q\to\infty$)).
After canceling numerators and denominators 
in suitable pairs, $S$ is estimated
as $-\ln\frac{1+\sin\frac{\pi m}{Q}}{1-\sin\frac{\pi m}{Q}}+\frac{r_m}{Q^{\beta_m}}
+o(\frac{1}{Q^{\beta_m}})$ 
where $Q^{\frac{2}{3}}\ll N\ll Q$ and 
$r_m$ and $\beta_m$ are constants
independent of $Q$.
Since we have two parameters $\frac{s}{Q^{\delta}}$ 
and $\frac{r_m}{Q^{\beta_m}}$,
we need another set of (similar) equations for the BA roots 
which can be derived
from the functional equation (\ref{feq}) \cite{kra}.
From the two sets of the equations, 
we can fix the parameters as $\delta=\beta_m=1$, $s=\pi(p+\frac{1}{2})$.
Therefore the asymptotic distribution of the $\theta_k$ 
is determined as (see Fig.\ref{fig:iqr})
\begin{eqnarray}
\theta_k=-\pi+\left(k+p+\frac{1}{2}\right)\frac{\pi}{Q}
+o\left(\frac{1}{Q}\right)
\quad(k\sim Q).\nonumber
\end{eqnarray}

\begin{center}
\begin{figure}
\epsfig{figure=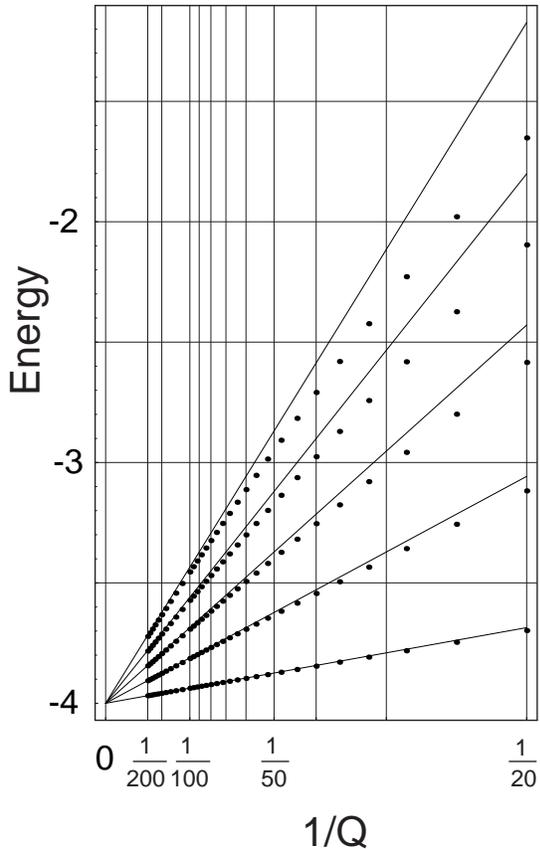,width=8.2cm}
\begin{minipage}{8.5cm}
\caption{\label{fig:iqe}
Numerical results for the energies of several bands
near the lower boundary.
Asymptotic behaviors are shown by the solid lines.
}
\end{minipage}
\end{figure}
\end{center}

Cases for the $2p$-th bands are also discussed in the similar way.
To summarize the results, 
the energy of the ($n+1$)th lowest band ($n=0,1,2,\ldots$;
$n=O(1)$) is given as
\begin{eqnarray}
\varepsilon_n
=-4+\left(n+\frac{1}{2}\right)\frac{4\pi}{Q}
+o\left(\frac{1}{Q}\right).\label{ll}
\end{eqnarray} 
In Fig.\ref{fig:iqe}, the five lowest energies 
are plotted as a function of $1/Q$ ($20\le Q\le200$)
and the straight lines are the asymptotic behavior in Eq.(\ref{ll}).

We can also treat the general energy bands 
near the higher boundary of the spectrum.
These ($n+1$)-th highest energies are given by
$\varepsilon_n^{\prime}=4-(n+\frac{1}{2})\frac{4\pi}{Q}+o(\frac{1}{Q})$ 
by the parallel argument.

Now let us discuss the physical outcome of the results. 
In the absence of the magnetic field, 
one can recover the parabolic dispersion 
in the continuum model from the tight binding model as 
\begin{eqnarray}
E_k&=&-2t(\cos k_xa+\cos k_ya)\cr
&&\stackrel{a\to0}{\to}-4t+ta^2k^2
=-4t+\frac{\hbar^2k^2}{2m^*}.\nonumber
\end{eqnarray}
It implies that the effective mass is given by $m^*=\hbar^2/2a^2t$. 
Since $1/Q=\phi=Ba^2/\Phi_0=|e|Ba^2/hc$, 
we rewrite the above result (\ref{ll}) with this effective mass $m^*$ as 
\begin{eqnarray}
E_n=t\varepsilon_n
\to-\frac{2\hbar^2}{m^*}\frac{1}{a^2}
+\hbar\omega_c\left(n+\frac{1}{2}\right)\quad(a\to0),\cr
n=0,1,2,\ldots;\,n=O(1),\nonumber
\end{eqnarray}
where $\omega_c=|e|B/m^*c$.
These are usual energies of the Landau levels except the diverging
energy shift. 
Now we can analytically recover the Landau levels from the BA
equations of the Bloch electrons in a magnetic field \cite{ld}.

We thank Y. Morita and K. Kusakabe for fruitful discussions. 
Y.H. was supported in part by a Grant-in-Aid from the Ministry of
Education, Science, and Culture of Japan. 
The computation in this work has been partly done 
at the YITP Computing Facility 
and at the Supercomputer Center, ISSP, 
University of Tokyo.

\end{multicols}

\begin{references}
\bibitem{del}
For the $k$ of the order $Q$, 
one can estimate $\Delta_k$ as
$\Delta_k$
$=(\Delta_k/\Delta_{k-1})$
$\cdots$
$(\Delta_{k/2+1}$
$/$
$\Delta_{k/2})$
$\Delta_{k/2}$
$\le$
$(\Delta_{k/2+1}/\Delta_{k/2})^{k/2}$
$\Delta_{k/2}$
$\le$
$C\chi(\theta^*)^{-\alpha Q}$
where $\theta^*$ and $\alpha$ are constants independent of $Q$ 
such that $\chi(\theta^*)<1$ and $0<\alpha<1$ \cite{kra}.
Therefore,
$\Delta_k$ is exponentially small
for the $k$ of the order $Q$.
\bibitem{ld}
To show the Landau degeneracy, we have to show that each of the bandwidth
vanishes in the semiclassical limit even after the rescaling
$\varepsilon_n\to E_n=t\varepsilon_n
=\frac{\hbar^2}{2m^*}\frac{1}{a^2}\varepsilon_n$
($a\to0$).
However the sum of the $Q$ energy bandwidths 
calculated numerically scales as $\sim Ct/Q$ \cite{tho2}. 
It implies that each bandwidth scales as $\sim
Ct/Q^2=C(\hbar^2/2m^*a^2)(|e|Ba^2/hc)^2\sim C^{\prime}a^2$. 
If we assume it, the bandwidth vanishes in the continuum limit ($a\to0$). 
To show the degeneracy analytically is an interesting future
problem.
\end{references}
\end{document}